 \documentclass[final,5p,times,twocolumn]{elsarticle}

\usepackage{amssymb}
\usepackage{lipsum}
\usepackage{orcidlink}

\journal{Physics Letters B}

\usepackage{amsmath}
\newcommand{\e}{E_{\text{Pl}}}

\begin{document}

\begin{frontmatter}

%% Title, authors and addresses

%% use the tnoteref command within \title for footnotes;
%% use the tnotetext command for theassociated footnote;
%% use the fnref command within \author or \affiliation for footnotes;
%% use the fntext command for theassociated footnote;
%% use the corref command within \author for corresponding author footnotes;
%% use the cortext command for theassociated footnote;
%% use the ead command for the email address,
%% and the form \ead[url] for the home page:
%% \title{Title\tnoteref{label1}}
%% \tnotetext[label1]{}
%% \author{Name\corref{cor1}\fnref{label2}}
%% \ead{email address}
%% \ead[url]{home page}
%% \fntext[label2]{}
%% \cortext[cor1]{}
%% \affiliation{organization={},
%%            addressline={}, 
%%            city={},
%%            postcode={}, 
%%            state={},
%%            country={}}
%% \fntext[label3]{}

\title{Experimental Bounds on Deformed Muon Lifetime Dilation}

%% use optional labels to link authors explicitly to addresses:
%% \author[label1,label2]{}
%% \affiliation[label1]{organization={},
%%             addressline={},
%%             city={},
%%             postcode={},
%%             state={},
%%             country={}}
%%
%% \affiliation[label2]{organization={},
%%             addressline={},
%%             city={},
%%             postcode={},
%%             state={},
%%             country={}}

\author[first]{Iarley P. Lobo\,\orcidlink{0000-0002-1055-407X}\,}
\ead{lobofisica@gmail.com}

\affiliation[first]{organization={Department of Chemistry and Physics, Federal University of Paraíba},%Department and Organization
            addressline={Rodovia BR 079 - km 12}, 
            city={Areia},
            postcode={58397-000}, 
            state={PB},
            country={Brazil}}

\author[second]{Christian Pfeifer\,\orcidlink{0000-0002-1712-6860}\,}
\ead{christian.pfeifer@zarm.uni-bremen.de}
\affiliation[second]{organization={ZARM, University of Bremen},%Department and Organization
            addressline={}, 
            city={Bremen},
            postcode={28359}, 
            state={},
            country={Germany}}

\author[third]{Pedro H. Morais\orcidlink{0000-0002-2226-3579}\,}
\ead{phm@academico.ufpb.br}
\affiliation[third]{organization={Physics Department, Federal University of Paraíba},%Department and Organization
            addressline={}, 
            city={João Pessoa},
            postcode={58059-900}, 
            state={PB},
            country={Brazil}}

\begin{abstract}
We analyze Planck scale induced modifications of the relativistic time dilation using data from the Muon Storage Ring experiment at CERN. By examining the lifetimes of muons, we establish, for the first time, a constraint on such quantum gravity-inspired deformations using this channel. The magnitude of the effect indicates that the study of cosmic rays is a well suited arena for this scenario. We show, that the spectrum of muons would be significantly affected for particles at the PeV scale. Since this later observation of the effect of time dilation is more indirect compared to a direct lifetime measurement, we encourage to perform a high precision measurement of the muon lifetime as a function of the muon's energy.
\end{abstract}

%%Graphical abstract
%\begin{graphicalabstract}
%\includegraphics{grabs}
%\end{graphicalabstract}

%%Research highlights
%\begin{highlights}
%\item Research highlight 1
%\item Research highlight 2
%\end{highlights}

\begin{keyword}
%% keywords here, in the form: keyword \sep keyword, up to a maximum of 6 keywords
Planck scale \sep Muon Lifetime \sep Quantum Gravity

%% PACS codes here, in the form: \PACS code \sep code

%% MSC codes here, in the form: \MSC code \sep code
%% or \MSC[2008] code \sep code (2000 is the default)

\end{keyword}

\end{frontmatter}

%\tableofcontents

%% \linenumbers

%% main text

\section{Introduction}
\label{introduction}

The debate over whether Poincaré symmetry is preserved at a fundamental level has been ongoing for decades, particularly in discussions about extensions of the Standard Model \cite{Kostelecky:2003fs} and quantum gravity phenomenology \cite{Amelino-Camelia2013,Addazi:2021xuf}. In general, the models discussed in the literature can be categorized into two scenarios: a) Lorentz Invariance Violation (LIV), where the relativity principle is broken and preferred inertial frames -- those in which the behavior of physical systems is distinguished -- exist; b) deformed (or doubly) special relativity (DSR), where a deformation or modification of Poincaré symmetry maps inertial frames onto each other and no distinguished preferred frame exists.

In the LIV case, one still has the map between frames that are related by relative velocities, translations and rotations as the standard Poincaré transformations, however, these are no longer symmetry transformations of the physical system described which thus leads to the existence of preferred frames. In the DSR case, the transformations between inertial frames are modified and adopted to the new symmetry of quantum spacetime. Usually, these modifications are studied as deviations from Poincar\'e symmetry by corrections suppressed by an energy scale (in the context of quantum gravity usually assumed to be the Planck scale) that indicates deviations from Poincaré symmetry \cite{Lukierski:1991pn,Lukierski:1992dt,Majid:1994cy,Amelinodsr1}. In general, both models predict some effects such as energy-dependent time delays in the time of arrival of simultaneously emitted massless particles, with corrections that are similar enough to make it challenging to distinguish between LIV and DSR models.

A series of papers \cite{Arzano:2019toz,Arzano:2020rzu,Bevilacqua:2022fbz,Bevilacqua:2024jpy,Lobo:2020qoa,Lobo:2021yem,Lobo:2023yvi,Morais:2023amp} showed that deformations of boosts in DSR present a promising avenue for investigating quantum gravity effects through the time dilation of particle lifetimes. In particular in \cite{Lobo:2020qoa,Lobo:2021yem,Lobo:2023yvi,Morais:2023amp} an amplification factor had been identified that is proportional to the square of the ratio between the energy and mass of the unstable particle $(E/m)^2$. If confirmed, this effect could serve as strong evidence that Poincaré symmetry needs to be extended at a fundamental level.

To place bounds on the modified time dilation effect and identify better ways to analyze this scenario with higher sensitivity, we examine one of the few cases for which the dilated lifetime of a fundamental particle is reported with a specific energy: the $g-2$ experiment performed at the Muon Storage Ring at CERN in \cite{Bailey:1977de,CERN-Mainz-Daresbury:1978ccd}. This paper represents the first instance of analyzing a boosted lifetime report from the perspective of a deformation of Lorentz symmetry by Planck scale effects, incorporating such a generic range of corrections. 

Since not much data is available for the direct comparison of lifetime measurements with time dilation predictions, we consider another, more indirect, observable, which is the muon flux on Earth generated by the interaction of cosmic rays with the atmosphere. We find and quantify the change of the muon flux influenced by modified time-dilations due to Planck scale MDRs.

This paper is organized as follows. In Section \ref{sec:time}, we describe how the deformed time dilation is computed in an effective spacetime. In section \ref{sec:bounds}, we analyze the data from CERN to set bounds on the deformed time dilation, and in section \ref{sec:LIVEarth} we investigate the impact of dilated muon lifetimes on the muon spectrum generated by cosmic rays interacting with the Earth's atmosphere. In section \ref{sec:discussion} we discuss our results. Throughout this article, we use units in which $\hbar=c=1$.

\section{Revisiting time dilation computation}\label{sec:time}
Our starting point is a generic class of modified dispersion relations (MDRs) that describe propagation through a quantum spacetime, expressed as:
\begin{equation}
 E^2-|p|^2=m^2+\xi^{(n)}\frac{|p|^{n+2}}{E_{\text{Pl}}^n}\,\label{eq:mdr1}\, ,
\end{equation}

where the sign of $\xi$, positive/negative, determines if the case considered is of superluminal/subluminal propagation, $E$ is the energy of the test particle, $|p|=|\vec{p}|$ is its spatial momentum and $m$ is its mass. We refer to $\xi^{(n)}$ as a dimensionless parameter that controls the kinematical deformations suppressed by the $n$-th power of the Planck energy $\e=\sqrt{\hbar c^5/G}\approx 1.22\times 10^{19}$ GeV. 
\par
A suitable spacetime geometry to describe the motion of such a particle through a quantum spacetime is Finsler geometry \cite{Girelli:2006fw,Pfeifer:2019wus}. The arc-length in Finsler geometry, derived from a Legendre transformation \cite{Rodrigues_2024} on the Helmholtz action of a free particle with the above MDR \eqref{eq:mdr1}, is shown to be, see \cite{Lobo:2020qoa,Morais:2023amp,Amelino-Camelia:2014rga,Pfeifer:2019wus,Raetzel:2010je,Letizia:2016lew,Lobo:2016xzq}:
\begin{equation}
    S=\int d\lambda F(x,\dot{x})=\int d\lambda \left[\sqrt{\dot{t}^2 - \dot{x}^2}+\frac{\xi^{(n)}}{2}\left(\frac{m}{\e}\right)^n\frac{|\dot{x}|^{n+2}}{(\dot{t}^2 - \dot{x}^2)^{\frac{n+1}{2}}}\right]\,. \label{eq:arclength} 
\end{equation}
Here $F$ is the so-called Finsler function and ``dot'' means derivative with respect to the generic parameter $\lambda$. The resulting particle trajectories extremising these length functions are called Finsler geodesics, and the deformed Poincaré transformations are isometries of the Finsler function. 

From these results, it is shown in \cite{Morais:2023amp} that the deformed Lorentz transformation connecting the rest frame of the particle and the laboratory (lab) frame can be derived from the extension of the clock postulate to Finsler geometry \cite{Lobo:2020qoa}. This means that the proper time $\tau$ measured by an observer comoving to a particle of mass $m$ is given by the arc-length $S$ of its trajectory in spacetime divided by the speed of light $c$. The energy and momentum of the particle in the lab frame is given by $p_{\mu}=m\partial F/\partial \dot{x}^{\mu} \sim (E,\vec p)$, see \cite{Lobo:2020qoa}, and we can derive the dilated lab frame time $t$ as

\begin{align}
    t_{\text{DSR}}&=\gamma_{\text{DSR}}\tau=\frac{E}{m}\left[1+\frac{n\xi^{(n)}}{2}\left(\frac{|p|}{m}\right)^{2}\left(\frac{|p|}{E_{\text{Pl}}}\right)^n\right]\tau\label{eq:dil_time1}\\
    &=\frac{\sqrt{|p|^2+m^2}}{m}\left\{1+\frac{n\xi^{(n)}}{\e^n}\frac{|p|^{2+n}\left[m^2(n+1)+np^2\right]}{2m^2(p^2+m^2)}\right\}\tau\, ,\nonumber
\end{align}
where in the last step, we used the MDR \eqref{eq:mdr1}. This result is a correction of the Lorentz transformation of time from the comoving frame to the lab frame. The verification that it is indeed a deformed Lorentz transformation was shown in \cite{Morais:2023amp}, where it was verified that this transformation of the time coordinated can be supplemented by transformations of the spatial coordinates such that the total transformation is a deformed Lorentz transformation, in the sense that it preserves the Finsler function (i.e.\ is an isometry of the Finsler space) as well as the original modified dispersion relation. A compatible modified energy-momentum conservation or 4-momentum addition law had also been constructed \cite{Lobo:2021yem}. How a compatible addition law for 4-velocities would need to look like has so far not been investigated and is an interesting question for future research.

We stress that such verification was confirmed at leading order in Planck scale correction, i.e., when satisfying the condition $\frac{n\xi^{(n)}}{2}\left(\frac{|p|}{m}\right)^{2}\left(\frac{|p|}{E_{\text{Pl}}}\right)^n\ll 1$.

In this letter, we deal with the phenomenological consequences of extending the clock postulate of Special Relativity to modified dispersion relations through the Finsler realm, by identifying the proper lifetime of an unstable particle in its comoving frame with its arc-length in spacetime. This expression gives the dilation of its proper lifetime $\tau$ by a deformed Lorentz factor, $\gamma_{\text{DSR}}$, which is consistent with a deformation of the Lorentz symmetry. This result suggests that such modification of the dilated lifetime is compatible with DSR principles. However, we stress that a complete analysis would require a computation of the lifetimes in the lab and comoving frames, considering possible deformations of the composition law of energy and momenta. Since this would require field theoretic tools based on Finsler geometry, that go beyond the scope of this letter, we limit ourselves to the proposal described in Eq.\eqref{eq:dil_time1}.

In the next section, section \ref{sec:bounds}, we will set bounds on the parameter $\xi^{(n)}$ using data from experiments measuring the anomalous magnetic moment of the muon $g-2$ at CERN. These experiments provide explicit values for the Lorentz factor and the dilated lifetime of the muon, allowing us to constrain $\xi^{(n)}$. Moreover, we will analyse the impact of the deformed dilated lifetime on muon-spectrum measured in the Earth in section \ref{sec:LIVEarth}. 

%%%%%%%%%%%%%%%%%%%%%%%%%%%%%%%%%%%%%%%%%%%%%%%%%%%%%%%%%%%%%%%%%%%%%%%%%%%%%%%%%%%%%%%%%%%%%%%%%%%%%%%%%%%%%%%%%%%%%%%%%%%%%%%%%%%%%%%%%%%%%%%%%%%%%%%%%%%%%%%%%%%%%%%%%%%%%%%%%%%%%%%%%%%%%%%%%%%%%%%%%%%%%%%%%%%%%%

\section{Bounds from the Muon Storage Ring experiment at CERN}\label{sec:bounds}
At the Muon Storage Ring at CERN, the anomalous magnetic moment of both muons and antimuons was measured, as detailed in the final report \cite{CERN-Mainz-Daresbury:1978ccd}. This experiment scrutinized the orbital and spin motion of highly polarized muons within a magnetic storage ring. The process involved proton collisions within a synchrotron accelerator, yielding pions that decayed into muons alongside neutrinos. Subsequently, these muons were injected into a region featuring a uniform magnetic field, where they were accelerated to traverse circular paths, thereby experiencing time dilation.
\par
Notably, the same procedure was employed in the recent $g-2$ experiment at the Fermilab \cite{Muong-2:2021ojo}. This endeavor significantly enhanced the precision of CERN's experiment. However, while both experiments aimed to measure the anomalous magnetic moment, the CERN setup also delved into quantifying the dilation of muon lifetimes—a facet not explored in Fermilab $g-2$, which primarily focused on measuring the muon's magnetic moment. Therefore, we have opted for the CERN experiment as it is more conducive to establishing constraints on a modified time dilation effect.
\par
In \cite{Bailey:1977de}, the reported momentum of the muon and the weighted average of the measured lifetime for both muons and antimuons, along with its mass and mean lifetime according to the Particle Data Group (PDG) \cite{ParticleDataGroup:2022pth}, are
\begin{equation}
\begin{aligned}
|p| &= 3.094\, \text{GeV}, & \qquad t_{\text{exp}} &= 64.378 \pm 0.026\, \mu\text{s}, \\
m_{\mu} &= 105.658\, \text{MeV}, & \qquad \tau_{\mu} &= 2.197\, \mu\text{s}.
\end{aligned}
\end{equation}
\par
Using these values, we can calculate the bounds on the quantum gravity parameter $\xi^{(n)}$ by considering the maximum and minimum values of $t_{\text{exp}}$. We insert this information in Eq.\eqref{eq:dil_time1}, and compare with the maximum and minimum experimental values to derive the bounds. For positive and negative $\xi$'s, the bounds for $n=1$ and $n=2$ are of the order
\begin{align}
&-2.940\times 10^{12}\leq\xi^{(1)}\leq 4.481\times 10^{12}\, , \\ &-5.780\times 10^{30} \leq \xi^{(2)}\leq 8.839\times 10^{30}
\end{align}

These bounds are far from the Planck scale, a consequence of the measurements being conducted with particles at GeV energy levels. This is typical of experiments aimed at measuring the anomalous magnetic moment of the muon. This value of the Lorentz factor effectively removes the contribution of the stabilizing quadrupole electrostatic field from the muon’s relation between the angular frequency and the electromagnetic field, described by the Thomas-Bargmann-Michel-Telegdi equation~\cite{jegerlehner2008anomalous}. 

With this observation, we would like to point out that a dedicated high-precision experiment measuring the muon-lifetime in the laboratory frame dependent on their energy, would be a formidable window to Planck scale physics. We could improve the bounds presented here easily if better data would be available.

Due to the absence of better data, we will search for consequences of the dilated lifetime in further observations, which are however, more indirect. Since this effect has an amplifier of the order $(|p|/m)^2$, see \eqref{eq:dil_time1}, improvements could be achieved by examining observables that depend on the dilated lifetime of more energetic particles, and for which data is available, such as the spectra of particles produced by cosmic rays \cite{Lobo:2021yem}.

%%%%%%%%%%%%%%%%%%%%%%%%%%%%%%%%%%%%%%%%%%%%%%%%%%%%%%%%%%%%%%%%%%%%%%%%%%%%%%%%%%%%%%%%%%%%%%%%%%%%%%%%%%%%%%%%%%%%%%%%%%%%%%%%%%%%%%%%%%%%%%%%%%%%%%%%%%%%%%%%%%%%%%%%%%%%%%%%%%%%%%%%%%%%%%%%%%%%%%%%%%%%%%%%%%%%%%
\section{Testing Lorentz Invariance Violation Using Cosmic Rays on Earth}\label{sec:LIVEarth}

The interaction of cosmic rays with the Earth's atmosphere provides a unique opportunity to test deviations from Lorentz invariance. The potential elongation of pion lifetimes due to a modified dispersion relation for neutrinos, leading to predicted changes in the energy spectra of neutrinos and muons has been discussed in \cite{Cowsik:2012qm}. This section builds on that work by analyzing the effects of potential modifications in the muon lifetime itself and examining the subsequent impact on the observed muon energy spectrum.

Following \cite{Cowsik:2012qm}, the rate of muon production via pion decay is expressed as (Eq.~(44)):
\begin{equation}
    q_{\mu}(E,x,\theta)=A_{\pi} x e^{-x/\lambda_{\pi}} E^{-(\gamma+1)}\eta_{\mu}^{\gamma}\left\{\frac{E/\eta_{\mu}}{\varepsilon_{\eta_{\mu}}+E/\eta_{\mu}}\right\}\frac{1}{x}\frac{\varepsilon_{\eta_{\mu}}}{E/\eta_{\mu}}\, ,\label{eq:rate}
\end{equation}
where $A_{\pi}$ is a normalization constant, $x$ is the column density that a cosmic ray particle must penetrate to reach a given point in the atmosphere, $E$ is the muon energy, $\lambda_{\pi} \approx 120$ g/cm\(^2\) is the interaction mean free path of pions, $\gamma \approx 1.7$ determines the power law dependence. Although the dynamics of pion decay should be modified in this scenario, we assume as an ansatz that the effective average fraction of energy transferred to the muon during pion decay does not differ significantly from the standard result, leading us to adopt $\eta_{\mu} \approx 0.75$ for this fraction.

The term that is most susceptible to modifications of the dilated muon lifetime is the critical energy:
\begin{align}\label{eq:critEN}
    \varepsilon_{\eta_{\mu}} 
    = \frac{h_0\sec(\theta)m_{\mu}c^2}{c t_{\mu}}
    = E \left(\frac{h_0\sec(\theta)}{c t_{\mu} \tfrac{ E }{m_\mu c^2} }\right)\,.
\end{align}
Here, $h_0=7\times 10^5$ cm is the atmospheric scale height, $\theta$ is the zenith angle of the incoming cosmic ray, $m_{\mu}\approx 105.7$ MeV is the muon mass, $c =3\times10^{10} \text{cm}/\text{s}$ is the speed of light and the muon lifetime in the laboratory frame $t_\mu$, appears explicitly. The usual Lorentz factor is already included in the terms $E$ of Eq.\eqref{eq:rate}, which leaves the term $t_{\mu}$ as, see \eqref{eq:dil_time1},
\begin{equation}
t_{\mu}=\left[1+\frac{n\xi^{(n)}}{2}\left(\frac{|p|}{m_\mu}\right)^{2}\left(\frac{|p|}{E_{\text{Pl}}}\right)^n\right]\tau_{\mu}\, ,
\end{equation}
where $p_{\eta}$ is the muon momentum and $\tau_{\mu}$ is the muon lifetime at rest. In fact, the critical energy \eqref{eq:critEN} is the key quantity that is modified when one analyzes the modified dilated lifetime, as discussed in \cite{Cowsik:2012qm}. Further DSR effects, which might influence the muon production, like modifications in energy thresholds of particle interactions and cross-sections, only add faint subdominant corrections to the expression, as has been demonstrated extensively in the literature  \cite{Amelino-Camelia:2010lsq,Lobo:2021yem,Carmona:2021lxr,Carmona:2023luz}. Thus, we neglect such contributions and focus on the effect of modified Lorentz transformations on the time dilation, as they yield the dominant correction in this analysis.

\begin{figure}[ht]
    \centering
\includegraphics[width=.48\textwidth]{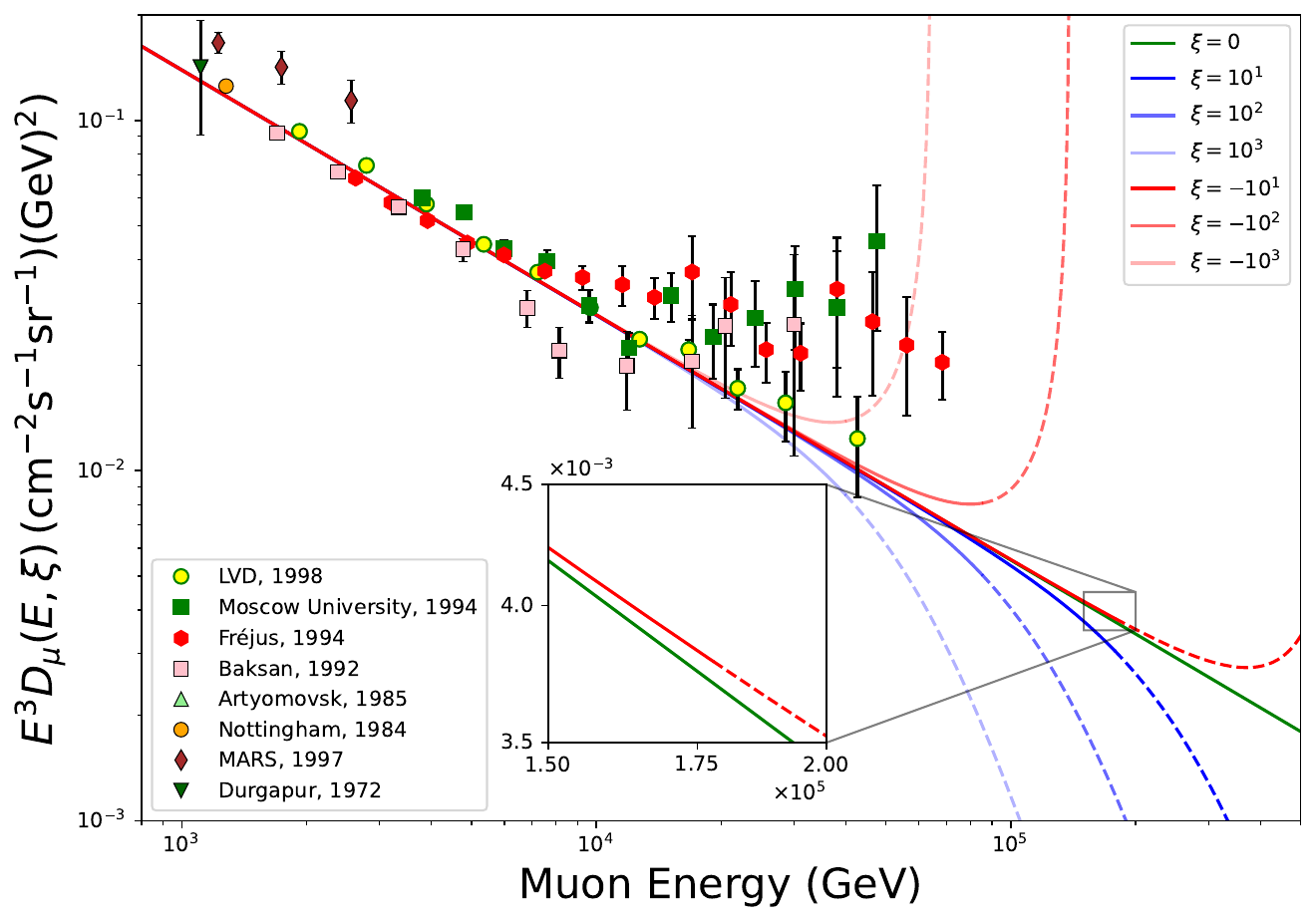}
    \caption{Vertical differential momentum spectrum of muons at sea level. The direct data are taken from Refs. \cite{Baber:1968sha,Bateman:1971cc,Allkofer:1971qr,Nandi:1972ns,Ayre:1975qi,Rastin:1984nu,DePascale:1993jgv,Bugaev:1998bi} and indirect (underground) data are from Refs. \cite{andreyev1990,ito1990,Miyake:1964ula,Bakatanov:1992gp}. The impact of quantum gravity induced time dilations are shown in red, for negative $\xi$ / subluminal propagation, and blue, for positive values of $\xi$ / superluminal propagation. The green curve represents the classical special relativity case. The values $\sec(\theta)=1$, $A_\pi = 5.66\times10^{14}$ and $n=1$ were used to produce this plot.}
    \label{fig:spectrum}
\end{figure}

The muon intensity at the Earth's surface, $D_{\mu}$, which is the main observable we are interested in, can be calculated by integrating the production rate \eqref{eq:rate} from 0 to $x_{max}=x_0\sec(\theta)$, where $x_0=1030$ g/cm\(^2\):
\begin{equation}\label{eq:spectrum}
    D_{\mu}(E,x_{max},\theta)=\int_0^{x_{max}}q_{\mu}(E,x,\theta)dx\, .
\end{equation}
This energy dependent so-called differential spectrum is compared with experimental data collected over the past decades. Figure~\ref{fig:spectrum} shows the spectrum as a function of muon energy alongside data from Ref.~\cite{Bugaev:1998bi}. At low energies, Eq.~\eqref{eq:spectrum} provides a good description of the experimental results. However, discrepancies become apparent around $10^4$~GeV. This deviation is typically attributed to different models of the production of particles containing charm quarks (charm production), see~\cite{Bugaev:1998bi}. In general, these models increase the spectrum at higher energies, bringing it closer to the experimental data.

We study and show in Fig.~\ref{fig:spectrum} how a quantum gravity induced modified dilated muon lifetime changes the predicted muon flux for different values of the parameter $\xi^{(1)}$ (which we simply call $\xi$) for $n=1$. The solid lines correspond to the description of $E^3 D_{\mu}$ considering that the dimensionless correction $\frac{n\xi^{(n)}}{2}\left(\frac{|p|}{m_\mu}\right)^{2}\left(\frac{|p|}{E_{\text{Pl}}}\right)^n \ll 1$ of our Finsler approach, to depict the regions in which departures start to appear. The dashed lines represent an extrapolation of the plot beyond the perturbative level. This extrapolation will most likely not hold when more terms in the power series expansion in $E_{\text{Pl}}$ are taken into account. The expectation is that for higher order expansion, the curves converge more towards the measured data points. In this sense our finding determines viable coefficients for a first order Taylor expansion, from which the series has to be continued to reproduce the observed data. In order to find the all order imprint in the muon intensity of the time dilation, an all order derivation of the Finsler arc-length from an all order MDR has to be performed, which is still a matter of investigation.

For negative $\xi$, corresponding to subluminal propagation, a value of $\xi\sim -10^{3}$ pushes the spectrum upwards, closer to the experimental data. For positive and negative values of $\xi$, the muon intensity \eqref{eq:spectrum} decreases, however the negative $\xi$ case gives a contribution that when described by a power law is not as strong as $E^{-3}$, which is the reason why the red curves are pushed upwards. Conversely, the case of positive $\xi$, corresponding to superluminal propagation, is strongly disfavored, because it contributes to a decrease in the spectrum, contradicting the trend observed in the experimental data. Although this result was found in an approach that deforms Lorentz symmetry, it is consistent with other studies, based on Lorentz Invariance Violation, that impose stringent constraints on the superluminal scenario, like \cite{HAWC:2019gui}.

We stress that in both cases the rate of muon production decreases with the energy of the particles. The difference is that for negative $\xi$, this decrease in this rate is not strong enough to compete with the $E^3$ factor considered when studying the differential spectrum in Fig.\ref{fig:spectrum}, and this pushes the red curves upwards for higher energies. The opposite behavior happens for positive $\xi$.

Analyzing the contribution of the quantum gravity modification of the critical energy to the differential spectrum, as in Eq.~\eqref{eq:spectrum}, reveals that this quantity can approximately be expressed as  
\begin{equation}\label{eq:Ddsr}
D_{\mu}^{\text{DSR}} \approx D_{\mu}^{\text{SR}} \left(1 - \frac{n \xi^{(n)}}{2} \left(\frac{|p|}{m_\mu}\right)^2 \left(\frac{|p|}{E_{\text{Pl}}}\right)^n \right),
\end{equation}
where $D_{\mu}^{\text{SR}}$ represents the standard spectrum derived using special relativity. It is evident that a positive (negative) $\xi$ lowers (raises) the spectrum for particles with a larger momentum.

This behavior arises due to modifications in the time dilation effect described in Eq.~\eqref{eq:dil_time1}, where a positive (negative) $\xi$ increases (decreases) the muon's lifetime in the laboratory frame, allowing it to travel farther before decaying.

\begin{figure}[htp]
\centering
\includegraphics[width=.48\textwidth]{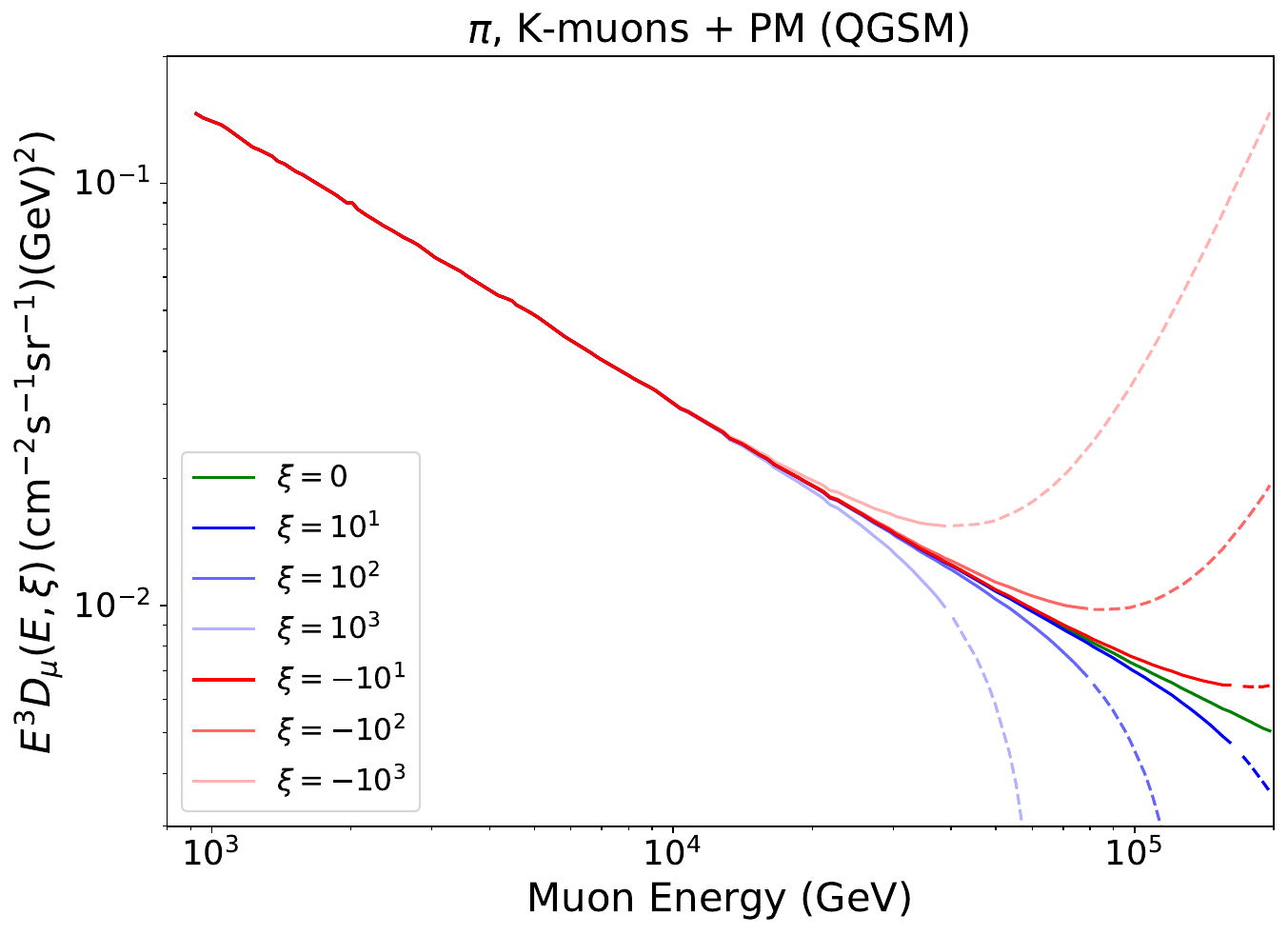}\\
\includegraphics[width=.48\textwidth]{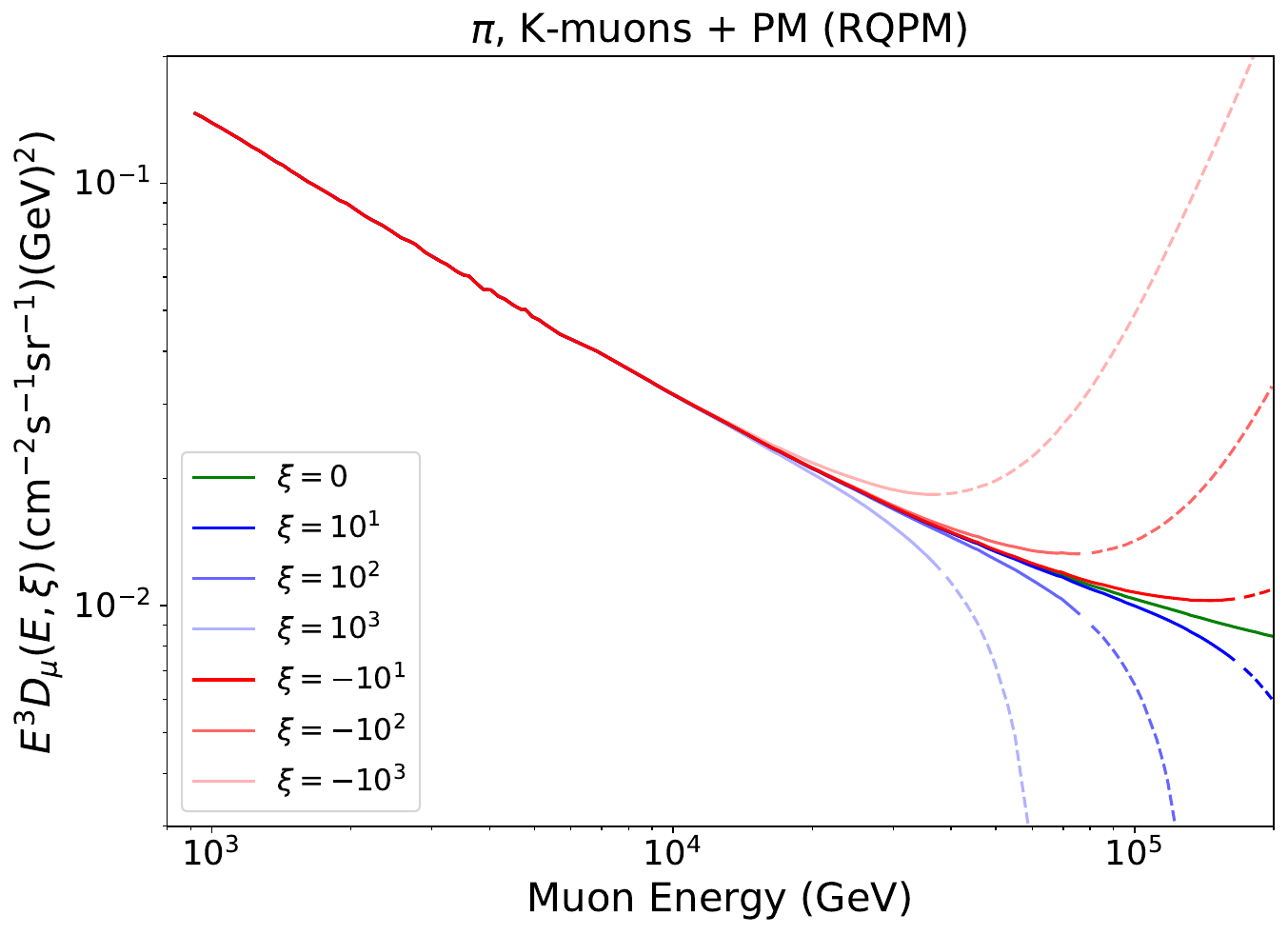}\\
\includegraphics[width=.48\textwidth]{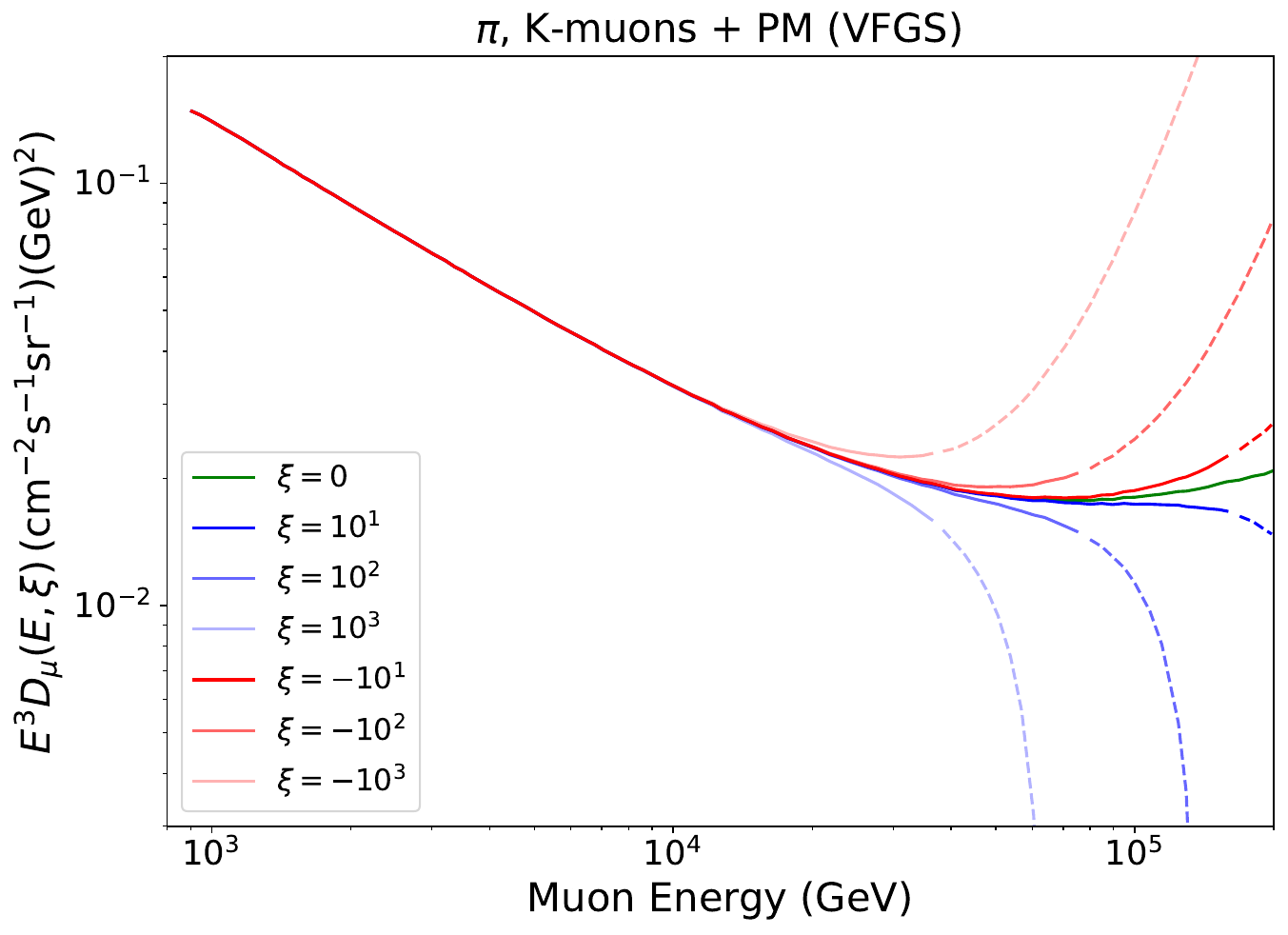}
\caption{Deformation of the differential spectrum considering as basis different models of charm production. On the top, QGSM, in the middle RQPM, and on the bottom VFGS model. We are using the same parameters used in Fig.~\ref{fig:spectrum}.}
\label{fig:charm}
\end{figure}

Next, we use ~\eqref{eq:Ddsr} to demonstrate that this general trend persists when considering more refined models that include additional effects, such as the charm production, as $\xi=0$ background. We follow ~\cite{Bugaev:1998bi} and consider the Quark-Gluon String Model (QGSM), the Recombination Quark Parton Model (RQPM) \cite{Bugaev:1989we,Bugaev:1987dm,Bugaev:1989tj}, and the semiempirical model by Volkova {\it et al.} (VFGS) \cite{Volkova:1987gh}. The resulting spectra are displayed in Fig.~\ref{fig:charm}.

We find that a negative $\xi$ amplifies the spectra, whereas a positive $\xi$ reduces it at higher energies. The latter case is particularly intriguing for phenomenological studies, as it represents a significant shift in the predictions of the background models. This behavior serves as a potential signature of superluminal quantum gravity effects, providing the means to impose even more stringent constraints on this scenario, as they already exist.

Another intriguing observation is that effects of order $\xi \sim 1$--$10$ become detectable at energies around the $100$~TeV scale. This makes muons prime candidates for probing deviations from Lorentz invariance induced by Planck-scale effects \cite{Bogdanov_2012}. As shown in Fig.~\ref{fig:charm}, quantum gravity effects can compete with, and even mimic, those arising from various particle physics models that exhibit similar deformations at the $100$~TeV scale.

%%%%%%%%%%%%%%%%%%%%%%%%%%%%%%%%%%%%%%%%%%%%%%%%%%%%%%%%%%%%%%%%%%%%%%%%%%%%%%%%%%%%%%%%%%%%%%%%%%%%%%%%%%%%%%%%%%%%%%%%%%%%%%%%%%%%%%%%%%%%%%%%%%%%%%%%%%%%%%%%%%%%%%%%%%%%%%%%%%%%%%%%%%%%%%%%%%%%%%%%%%%%%%%%%%%%%%

\section{Discussion}\label{sec:discussion}

We have calculated the first direct bound on a modification of time dilation due to DSR effects, from the lifetime of elementary particles. This effect can be derived by extending the clock postulate to an effective description of the propagation of particles and fields on quantum spacetime through a Finsler spacetime. It incorporates quantum gravity induced modifications into particle kinematics geometrically. This approach and its resulting predictions are compatible with a deformation of special relativity, as the Finsler norm is preserved by such transformations, as shown in \cite{Morais:2023amp}. In Special Relativity, we know that such kinematical postulate is compatible with the analysis of the decay of particles in the comoving and lab frames. A dynamical analysis of this effect  requires a field theoretic formulation based on this Finsler framework. This area is currently under development \cite{Hohmann:2021zbt}, and we expect to explore more of this issue in the future.

We compared the lifetimes of particles in the comoving frame with those in the lab frame. To directly set a bound, we used an experiment where the momentum of a massive particle is known with high accuracy and the corresponding dilated lifetime was measured directly and published. The appropriate experiment was conducted at the CERN Muon Storage Ring \cite{CERN-Mainz-Daresbury:1978ccd}, which provided data on measurements of relativistic time dilation in a controlled environment \cite{Bailey:1977de}.

We found that the bound for the first-order correction lies 12 orders of magnitude above the Planck scale, and for the second-order correction, it lies 30 orders of magnitude above. The smallness of the effect is due to the relatively low energies of the muons, approximately in the GeV range. To improve these constraints, it would be necessary to either consider experiments or observations with particles at higher energies or to enhance the precision of the measurements. Dedicated experiments which directly measure the lifetime of elementary particles, such as muons, are thus a perfect opportunity to probe Planck scale physics. Due to nowadays absence of such dedicated experiments, we also consider observables that include the dilated lifetimes of particles more indirectly to probe the effect. 

We showed that the spectrum of detected muons could be affected due to dilated lifetimes at scales that are accessible with current observations. The spectra are raised for negative dimensionless quantum gravity parameters (subluminal propagation of massless particles) and decreased for positive ones (superluminal propagation of massless particles). For Planck scale corrections, at first order, such departures become relevant at energies around $10^{5}-10^{6}$ GeV. The data considered in this paper goes at most to a region around $10^4$ GeV, which means that an improvement in one or two orders of magnitude in the measurement of the spectrum, would allow one to scrutinize deformations of time dilation with Planck scale sensitivity. In order to try to describe the data for energies lower than $10^5$ GeV only with quantum gravity, it would be necessary to have corrections $10^3$ times stronger than typical Planck-scale effects and to go beyond the perturbative level at first order. Nevertheless, the analyses described in this paper show the potential of a phenomenological approach based on the modified time dilation for the near future. We also stress that constraints on the quantum gravity scale based on violations of the relativity principle do not apply to our model, neither those that modify massless particles' equations.

In conclusion, the lack of data for direct measurements of muon lifetimes as a function of the energy in the lab frame limits the direct constraints, which can be put on Planck-scale MDRs from time dilations. A dedicated experiment, for example in a muon accelerator device \cite{Lobo:2023yvi}, is very desirable to test Planck-scale physics.

%%%%%%%%%%%%%%%%%%%%%%%%%%%%%%%%%%%%%%%%%%%%%%%%%%%%%%%%%%%%%%%%%%%%%%%%%%%%%%%%%%%%%%%%%%%%%%%%%%%%%%%%%%%%%%%%%%%%%%%%%%%%%%

\section*{Acknowledgments}
I. P. L. was partially supported by the National Council for Scientific and Technological Development - CNPq grant 312547/2023-4. CP acknowledges support by the excellence cluster QuantumFrontiers of the German Research Foundation (Deutsche Forschungsgemeinschaft, DFG) under Germany's Excellence Strategy -- EXC-2123 QuantumFrontiers -- 390837967 and was funded by the Deutsche Forschungsgemeinschaft (DFG, German Research Foundation) - Project Number 420243324. P. H. M. thanks Coordena\c c\~ao de Aperfei\c coamento de Pessoal de N\'ivel Superior - Brazil (CAPES) - Finance Code 001 for financial support. The authors would like to acknowledge networking support by the COST Action BridgeQG (CA23130), supported by COST (European Cooperation in Science and Technology).

\bibliographystyle{elsarticle-num} 
\bibliography{references}
\end{document}